\documentclass[preprint2]{aastex}
\usepackage{aastexug}
\newcommand{\pks}{PKS~0637--752}
\usepackage{graphicx}
\usepackage{fancyheadings}
\usepackage{color}
\begin{document}

\title{Chandra Discovery of a 100 kpc X-ray Jet in PKS~0637--752 }
\author{D.A. Schwartz\altaffilmark{1},
H.L. Marshall\altaffilmark{2},
J.E.J. Lovell\altaffilmark{3},B.G. Piner\altaffilmark{4},S.J. Tingay\altaffilmark{3,4},
M. Birkinshaw\altaffilmark{1,5},G.
Chartas\altaffilmark{6},M. Elvis\altaffilmark{1}, 
E.D. Feigelson\altaffilmark{6}, K.K. Ghosh\altaffilmark{7},
D.E. Harris\altaffilmark{1},H. Hirabayashi\altaffilmark{8},
E.J. Hooper\altaffilmark{1}, D.L. Jauncey\altaffilmark{3}, K.M. Lanzetta\altaffilmark{9},  S. Mathur\altaffilmark{10},
 R.A. Preston\altaffilmark{4},  
W.H. Tucker\altaffilmark{1}, S. Virani \altaffilmark{1}, B. Wilkes\altaffilmark{1} 
and D. M. Worrall\altaffilmark{1,5} }
\altaffiltext{1}{Harvard-Smithsonian Center for Astrophysics} 
\altaffiltext{2}{Massachusetts Institute of Technology} 
\altaffiltext{3}{ATNF/CSIRO} 
\altaffiltext{4}{Jet Propulsion Laboratory} 
\altaffiltext{5}{University of Bristol} 
\altaffiltext{6}{Pennsylvania State University} 
\altaffiltext{7}{NASA/Marshall Space Flight Center} 
\altaffiltext{8}{Institute of Space and Astronautical Science} 
\altaffiltext{9}{State University of New York at Stony Brook} 
\altaffiltext{10}{Ohio State University}

\email{pks0637@head-cfa.harvard.edu}
\slugcomment{DRAFT, \today, das}

\begin{abstract}
The quasar \pks, the first celestial X-ray target of the Chandra X-ray
Observatory, has revealed asymmetric X-ray structure extending from 3
to 12 arcsec west of the quasar, coincident with the inner portion of
the jet previously detected in a 4.8 GHz radio image \citep{tin98}.
At a redshift of $z=0.651$, the jet is the largest ($\sim 100$ kpc)
and most luminous ($\sim 10^{44.6}$ ergs s$^{-1}$) of the few so far
detected in X-rays.  This letter presents a high resolution X-ray
image of the jet, from 42 ks of data when \pks\ was on-axis and ACIS-S
was near the optimum focus.  For the inner portion of the radio jet,
the X-ray morphology closely matches that of new ATCA radio images at
4.8 and 8.6 GHz. Observations of the parsec scale core using the VSOP
space VLBI mission show structure aligned with the X-ray jet, 
placing important constraints on the X-ray source models. HST images
show that there are three small knots coincident with the peak radio
and X-ray emission. Two of these are resolved, which we use to
estimate the sizes of the X-ray and radio knots. The outer portion of
the radio jet, and a radio component to the east, show no X-ray
emission to a limit of about 100 times lower flux.

The X-ray emission is difficult to explain with models that
successfully account for extra-nuclear X-ray/radio structures in other
active galaxies.  We think the most plausible is a synchrotron self-Compton
(SSC) model, but this would imply extreme departures from the
conventional minimum-energy and/or homogeneity assumptions. We also
rule out synchrotron or thermal bremsstrahlung models for the jet
X-rays, unless multicomponent or ad hoc geometries are invoked.

\end{abstract}


\section{Introduction}

\pks\ was the first celestial X-ray target of the Chandra X-ray
Observatory \citep{weiss00}. As a moderate strength point source it
was used to locate the optical axis and focus of the X-ray mirror
assembly. Surprisingly, even the first short, out-of-focus image
clearly revealed an X-ray jet, coincident with the radio jet reported
by \cite{tin98}. This letter addresses the X-ray jet, which appears as
a extension from 3 to $11\farcs5$ west of the quasar, with brighter 
condensations from $7\farcs5$ to $9\farcs5$.  This
corresponds\footnote{We use $H_0 = \rm 50\, km\, s^{-1}\, Mpc^{-1}$
and $q_0 = 0$ throughout} to an extension in the plane of the sky from
$\sim30$ to $\sim100$ kpc from the nucleus, with an X-ray luminosity
of $\sim ~ 4.2 \times 10^{44}$ erg s$^{-1}$: the largest and most
luminous X-ray jet discovered to date. In a companion paper
\citep{cha00} we examine the X-ray spectra of the core and jet in more
detail by adding substantial additional data where the detector was
slightly out of focus.

\pks\ was identified with the stellar object by \citet{hun71},
based on an accurate radio position, and a redshift of z = 0.651 was
measured by \citet{sav76}.
The HEAO-1 all sky survey
suggested it as a $2-10$ keV X-ray source \citep{wood84}. Definitive X-ray
identification was made in the $0.3-3.5$ keV band with the \emph{Einstein} observatory
\citep{elvis84}, and numerous X-ray observations have since been made
(cf. \citet{yaqoob98}, and references therein), however none
approached the sub-arcsecond image quality of Chandra \citep{jerius00}
and so could not resolve the jet. The source is
gamma-ray quiet \citep{fic94}, with a 2$\sigma$ upper limit of $4
\times  10^{-8}\, \rm photons \, cm^{-2} \, s^{-1}$ for gamma-rays
above 100 MeV.


VLBI space observatory program (VSOP) observations were rescheduled to overlap
the Chandra observations to investigate links between the known
pc-scale jet \citep{tin98} and the X-ray emission from
the quasar core. During the VSOP observations we also obtained
Australian Telescope Compact Array (ATCA) radio images with comparable resolution to the Chandra
images. We discuss the radio data in section \ref{sec:radio}.

HST WFPC-2 images
were obtained fortuitously on 30 May 1999, and are discussed in section \ref{sec:optical}. 

\section{OBSERVATIONS}

\subsection{X-ray}
\label{sec:xray}
Six Chandra observations, totaling 42 ks  when
the target was on axis and the detector was within 0.25 mm of focus, were used to create an
X-ray image. We use the full energy band of the back side illuminated
S3 CCD chip, $\sim0.3-10$ keV.

The PSF of the resultant core image has a half-power radius of
$0\farcs42$ and a $90\%$ encircled energy radius of $1\farcs55$. The
image quality of the quasar core is broadened slightly by event
pile-up, indicated by over-representation of the high grades, as well as
the high counting rates of $\sim$ 0.4 \, s$^{-1}$.  In the direction
perpendicular to the jet, the core fits a Gaussian profile with an rms
of 0\farcs37.  The jet is faint enough that pile-up is not significant
at any point.

\begin{figure}[h]
  \begin{minipage}[c]{0.5\textwidth}
    \includegraphics*[width=3.7in]{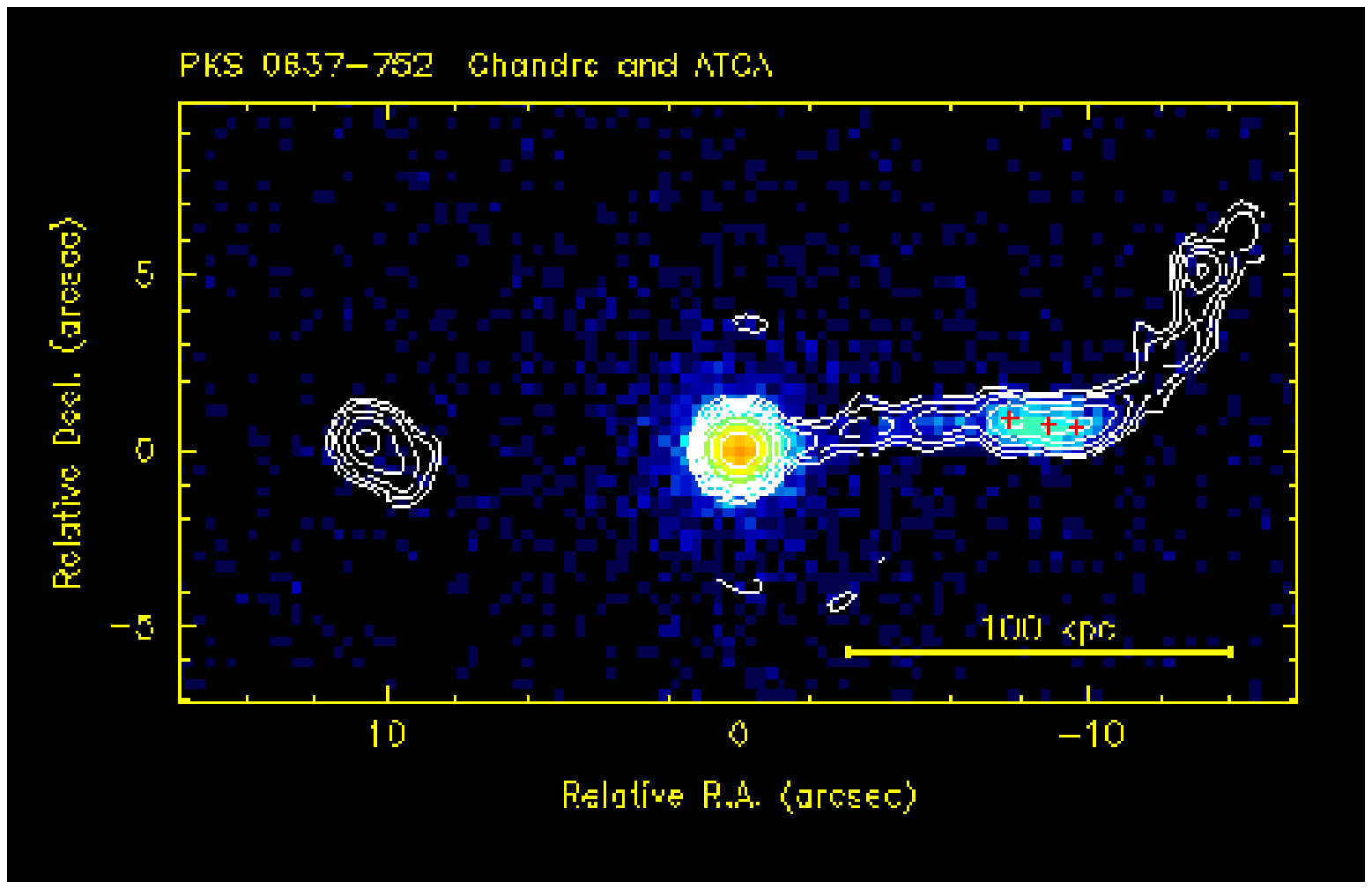}

	\hspace{0.6in}\includegraphics*[width=3.3in]{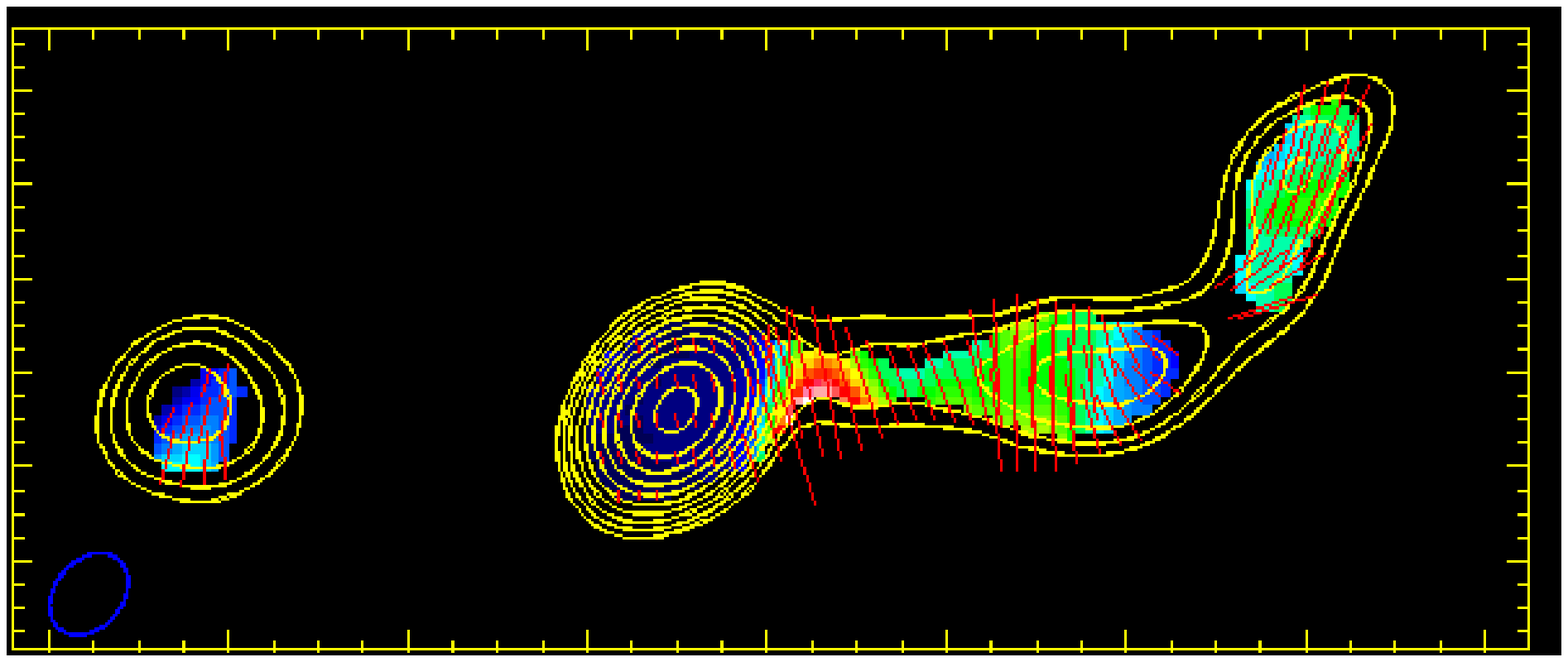} 
 \caption{\label{fig:images}
Top panel gives images of the quasar \pks\ in the X-ray (false
 color), radio (contours), and optical (plus sign) bands. The 8.6 GHz beam is 0.80 x 0.97 arcsec FWHM, PA
of major axis is -46 degrees. The lower panel shows the radio contours
 at 4.8 GHz, the fractional polarization in false color
 (blue $\sim$ 5\%, green $\sim$ 15\%, red $\sim$ 25\%), 
with the direction of the
 electric vector given by the red lines. The bend in the radio jet,
 the change in radio polarization fraction and direction, and the
 termination of the X-ray emission are all coincident, 11\farcs5 west
 of the quasar.}
  \end{minipage}%
\end{figure}

The X-ray and radio images are shown in Figure~\ref{fig:images}.  The
X-ray jet, which corresponds to the inner portion of the western radio
jet before the bend to the northwest, separates into two regions:
within 7\arcsec\ of the core the X-ray flux is somewhat faint and
maintains a position angle (PA) of -82\arcdeg; at about 7\arcsec\ from
the core the jet bends distinctly to the south, and the flux is about
three times greater.  The increased brightening may be dominated by
three point-like knots, as detected with HST, and shown as the plus
signs. In the brighter parts of the jet, a fit to the cross-jet
profile is consistent with the core PSF, thus the jet is unresolved in
this direction, with an intrinsic width less than 0\farcs3 FWHM.

We detect a net 1205 X-ray counts from the entire inner, western jet,
defined as a rectangle from 3\arcsec\ to 11\farcs5 west and from
0\farcs5 south to 2\arcsec\ north of the quasar. For the power law
energy index\footnote{We use $f_{\nu} \propto \nu^{-\alpha}$} of
$\alpha = 0.85 \pm 0.08$ \citep{cha00}, this corresponds to a
measured $2-10$ keV flux of $1.2 \times 10^{-13}$ erg cm$^{-2}$
s$^{-1}$, or a spectral normalization at 1 keV of $5.9 \times
10^{-14}$ erg cm$^{-2}$ s$^{-1}$ keV$^{-1}$.  That is, the flux
density is 25 nJy at $2.4 \times \, 10^{17}$ Hz. We fixed the galactic
absorption at $\rm N_{H}=9\times 10^{20}  cm^{-2}$.

\begin{figure*}
  \begin{minipage}[c]{0.5\textwidth}
\rotatebox{90}{\resizebox*{2.7in}{!}{\includegraphics*{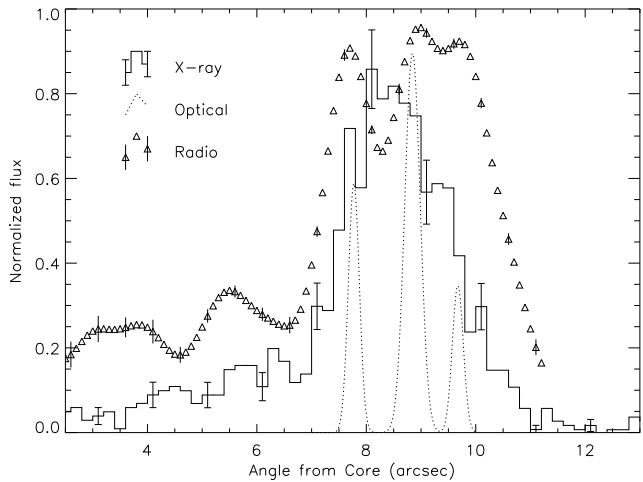}}}
\caption{\label{fig:profile} Radio and X-ray profiles of the large
scale jet in \pks.  The X-ray histogram gives counts in a 0\farcs2
bin, normalized to unity at 69 counts s$^{-1}$ bin$^{-1}$.  The
triangles are 8.6 GHz flux density per beam, normalized to unity at
0.04 Jy.  The dashed Gaussian curves mark the positions of knots
detected in the Hubble Space Telescope optical images, for an assumed
Gaussian profile.}
\end{minipage}
\end{figure*}

Figure~\ref{fig:profile} plots the X-ray and radio surface brightness
as a function of the distance West of the quasar core, integrated
$\pm$ 1\arcsec\ pperpendicular to the jet.  There are three distinct,
partially resolved radio peaks or knots in the brighter part of the
inner western jet, which we designate by their direction and distance
from the core using the notation: WK7.8, WK8.9, and WK9.7.  The
enhanced X-ray emission is closely associated with those knots, but
not in exact detail. In particular, the X-ray flux falls off more
steeply past 9\farcs5. The ratio of the flux densities at  1 keV  to
8.6 GHz  varies by no more
than a factor of 2 from the value $\rm 1.3 \times 10^{-7}$, in the
region $\sim4$ to $9.5$ arcsec west of the quasar core.


\subsection{Radio}
\label{sec:radio}

\pks\ is the subject of ongoing VSOP \citep{hir98}
 observations to monitor
its pc-scale evolution \citep{tin99}, with one of the 
VSOP observations being rescheduled to overlap the
first Chandra observation. ATCA observations were
scheduled in parallel with the VSOP observations to
provide 1\arcsec\ and 2\arcsec\ rresolution radio images at 8.6
and 4.8 GHz, respectively, to complement the Chandra images.

These ATCA images confirmed the remarkable coincidence of the radio
and X-ray jets. Subsequent observations were made to improve image
sensitivity and to determine the jet polarization.  The inner west
radio jet is optically thin with a spectral index of 0.81, and the
polarization is 10--20\%. The E-vectors are perpendicular
to the jet where X-rays are detected, but, as the X-ray flux decreases
near the bend in the radio jet,
the polarization position angle begins to change so that the E-vectors
become parallel to the jet's centerline for the remainder of the radio
jet.

We have reanalyzed our VLBI observations \citep{lovell00,cha00} 
to search for the presence of more compact components within the radio
knots. We find that the knots are indeed resolved at 0.05 arcsecond
resolution with no more than 5 mJy remaining at this resolution. This
suggests that the radio knots are most likely low surface brightness
"hot spots."

A rotation-measure image was constructed from the polarization data at
the two frequencies.  Faraday rotation was detected in the quasar core
(RM=80 rad m$^{-2}$) but \emph{not} in the jet, with an upper limit of $\pm$
30 rad m$^{-2}$.  The absence of significant Faraday rotation implies
that the intrinsic magnetic field in the jet is perpendicular to the
observed E vectors in Figure~\ref{fig:profile}. That is, the magnetic
field is longitudinal where the X-ray emission is strong, and
perpendicular to the jet where we do not detect X-ray emission.


The VSOP observations show that \pks\ displays apparent superluminal
motion in its parsec-scale jet, coaligned with the inner portion of
the arcsec-scale radio jet \citep{lovell00}.  Six datasets covering
1995--1999 (four from the U.S. Naval Observatory's geodetic VLBI
database \citep{pin98}, two from VSOP) allow the motion of three parsec-scale
features to be measured.  Linear least-squares fits to the separation
of these features from the VLBI core vs. time yield proper motions of
0.41$\pm$0.03, 0.29$\pm$0.05, and 0.36$\pm$0.09 milliarcseconds per
year from the outermost component inward \citep{lovell00}.  These proper motions and
associated errors are just consistent with all three components moving
at the weighted average proper motion of 0.36$\pm$0.02 mas\,yr$^{-1}$,
which corresponds to an apparent speed of (17.8)$\pm$(1.0)~c.  Since
the apparent transverse speed is given by $\beta_{obs} = \beta \, \sin
\theta /(1-\beta \, \cos \theta)$,\citep{rees66}, this apparent speed
places limits on the bulk Lorentz factor in the VLBI jet and the angle
of the VLBI jet to the line-of-sight of $\Gamma>$17.8 and
$\theta<$6$\fdg$4, respectively.  Since the parsec-scale and
arcsec-scale radio jets are very well aligned, unless the jet goes
through a large bend in a plane perpendicular to the plane of sky, the
actual length of the X-ray jet would be at least 940 kpc.

\subsection{Optical}
\label{sec:optical}

An optical image of \pks\ was obtained with the Hubble Space Telescope
WFPC2 using the F702W filter.  Three observations were combined in
order to eliminate cosmic rays, giving a total exposure time of 2100~s.  Three distinct knots are detected within the radio image contours,
7\farcs7, 8\farcs8, and 9\farcs6 west of the core
(Fig.~\ref{fig:images}). These have fluxes 0.2, 0.28, and 0.094
$\mu Jy$, respectively, at an effective frequency $4.3 \times 10^{14}$
Hz (6969 \AA). We have assumed an energy spectrum $f_{\nu} \varpropto
\, \nu^{-1}$ to deduce the flux densities and effective frequency. The knot at
7\farcs7 is not resolved, while the other two are about 0\farcs3 in
diameter.

The quasar image is saturated, so it is difficult to detect sources
within about 2\arcsec\ of the quasar core. We estimate that the knots
can be located relative to the quasar to an accuracy of about 0\farcs07.
The HST image also shows
the presence of a group of faint galaxies, $\sim$100 kpc in radius,
surrounding the quasar.

\section{DISCUSSION}

The strong polarization shows that the radio emission arises from
synchrotron radiation. Standard models assume a power law population
of relativistic electrons with a density $\rm n(\gamma)=n_0\,
\gamma^{-m}$ where $\gamma$ is the Lorentz factor. The radio spectral
index $\rm \alpha = (m-1)/2$ is observed to be 0.81, so that m=2.62.
A natural hypothesis is that the X-rays are also synchrotron radiation
from the same population of electrons. However, the optical flux falls
a factor of 10 below such a continuous spectrum, and therefore rules
out such a simple model. The absence of optical emission at the level
of $\rm 5.6\, \mu Jy$ which would be required, implies that the high
energy cutoff to the electron spectrum be such as to cause the radio
spectrum to steepen at  $\rm \nu < 3\times10^{12}$ Hz. For the
X-rays to result from synchrotron emission there would have to be an
independent population of electrons with a similar index m=2.7 in the
energy region emitting the X-rays, but which flattened at lower
$\gamma$ to avoid over-producing the optical. There then would be no
apparent reason for the existing spatial correlation of the radio and
X-ray emission.

Although the data allow a fit to a thermal spectrum with kT at least 4
keV \citep{cha00}, thermal bremsstrahlung is not a viable origin for
the X-ray emission unless a contrived geometry is invoked. Taking an
upper limit size of 0\farcs4 for the diameter of a cylindrical jet,
it would require an electron density of $\rm n_e = 2\,cm^{-3}$ to produce
the measured luminosity. But the upper limit to the rotation measure
places a limit $\rm n_e < 3.7\times10^{-5}/(H\,L)$. Even if the magnetic
field were as low as 1 $\rm \mu$ Gauss, a path-length of only $\rm L = 20$ pc
through such a thermal plasma would exceed the rotation measure limit.

The obvious remaining mechanism is inverse-Compton.  Because we have
measured the X-ray spectral index to be nearly the same as the 4.8 to
8.6 GHz radio index, it is natural to assume both arise from the same
population of relativistic electrons.  In inverse Compton scattering
scenarios the synchrotron radio emission is typically produced by the
higher energy electrons while the X-rays are produced by lower energy
electrons.  It is then natural to have high energy and low energy
cutoffs such that the optical emission is no larger than observed. 
In particular, we expect electrons of $\rm \gamma \sim 10^{4}$ to
scatter radio seed photons at $\sim$ 10 GHz up to the Chandra X-ray
energy range. The second order scattering would already be limited by
the Klein-Nishina cross section, and we would have no ``Compton
Catastrophe.'' 

To estimate the expected X-ray flux, we first apply standard
synchrotron theory, e.g. \citet{miley80}, to estimate the magnetic
field H and particle density $\rm n_0$ giving a minimum total
energy. For example, if WK7.8 is a sphere of 0.3 arcsec diameter,
uniformly filled with magnetic field and electrons, and with no proton
component, then H $\simeq$ 320 $\rm \mu$ Gauss, and the total minimum
energy in that sphere is $\rm U_{min} \simeq 1.5 \times 10^{57}$
ergs. We have assumed the radio spectrum extends from 10 MHz up to 500
GHz. In this situation, the predicted X-ray flux is $\sim 300$ times
less than observed. Such a model would satisfactorily explain the lack
of X-ray emission beyond the point where the radio jet bends toward
the northwest.

The radio synchrotron photons would have an energy density $8\times
10^{-11}$ ergs cm$^{-3}$ in the knot, if we assume the radio spectrum
extends up to 500 GHz. By comparison, the minimum energy magnetic
field calculated above would have an energy density $H^2/(8\pi) =
3.8\times 10^{-9}$ ergs cm$^{-3}$. In order that there be 300 times
more relativistic electrons, and predict the same GHz radio flux, we
must assume the magnetic field is $\rm \sim 6 \mu$Gauss.  To balance
the radio to X-ray flux ratios from the entire inner Western jet to
within a factor of 2 as is observed then requires the apparently
smooth jet to be composed primarily of many radio knots, with the
particle and field densities delicately balanced to produce the
X-rays. Such a gross departure from equipartition, increasing the
total energy by a factor of $\sim$ 1000, would pose significant problems
for models of  the particle acceleration and the jet
confinement.

We can make any inverse-Compton scenario more realistic by assuming
that the magnetic field strength need not be constant throughout the
volume. The exact distribution of field strength with volume would be
non-unique.  We must assume that the relativistic particles throughout
the entire volume produce the X-rays, while a small fraction of the
jet containing high magnetic fields provides essentially all of the
radio emission.

Alternate inverse Compton models could use some source of unseen
photons.  The equivalent luminosity of these photons in WK7.8 would
have to be $4\times10^{45}$ ergs s$^{-1}$.  If the core is the source
of this luminosity, it must radiate at an unreasonable value of $2
\times 10^{49}$ ergs s$^{-1}$, considering that the knot subtends a
solid angle of only $1.9 \times 10^{-4}$ sr.  We cannot rule out that
a Doppler-boosted beam is shining on the knots, but out of our direct
line of sight \citep{perez85}. This beam would presumably be
optical/IR emission. However, if this could happen then when we
\emph{did} fall in such a beam we would infer optical luminosities of
$\gtrsim 10^{49}$ ergs s$^{-1}$ for such a source. Such objects are
not observed.

We might invoke relativistic beaming of the jet and knots to reconcile
the observed X-ray flux with the apparent equipartition magnetic
field. The conventional formula \citep{jones74} for the ratio of SSC
X-ray to radio flux is multiplied by a factor $\delta^{-4-2\alpha}$
where $\delta = 1/(\Gamma (1-\beta\,\cos\theta))$ is the beaming
factor of matter moving with bulk Lorentz factor $\Gamma$ at an angle
$\theta$ to the line of sight (cf. \citet{mad83}). We can explain the
absence of X-rays from a putative eastern jet by Doppler suppression
if $\Gamma=8$, and then a value $\delta = 0.3$ would reconcile a
minimum energy radio source with the observed X-rays being produced by
SSC. This would require a bend of the VLBI jet, which must be at an
angle less than about 6 degrees, to an angle of order 53 degrees to
the line of sight.  The bend must be about an axis which is very
nearly in the plane of the sky, an unlikely coincidence.  Even with
such a coincidence, we note that the apparent radio luminosity of
$3\times10^{43}$ erg s$^{-1}$ would actually be a factor
$\delta^{2+\alpha} = 25$ higher in the rest frame, and for similar
sources beamed toward us we would infer a radio luminosity of
$2\times10^{46}$ erg s$^{-1}$. This  exceeds observed blazar
radio luminosities.

\section{CONCLUSIONS}

Discovery of the largest and most luminous X-ray jet in the very first
celestial X-ray target of Chandra has dramatically proven the power of
this observatory. It immediately shows the value of the \emph{two
dimensional} angular resolution improvement of a factor of 100 over
the best previous missions.  It is likely that many further X-ray jets
will be detected in extra-galactic radio sources. The X-rays place
difficult constraints on the physical conditions, eliminating standard
scenarios, and will surely have important astrophysical consequences,
e.g., in terms of understanding regions of particle acceleration,
inhomogeneities of magnetic field structures, and/or extreme
departures from equipartition conditions. While it seems that
inverse-Compton emission from the same electrons producing the
synchrotron radio emission is the most likely source of the X-ray
emission, we have noted difficulties with the specific scenarios we
have considered.

\acknowledgments 
This work was performed in part at the Jet Propulsion
 Laboratory, California Institute of Technology, under contract to
 NASA. We acknowledge NASA contracts and grants NAS 8-38252 to PSU,
NAS8-39073 to the Chandra X-ray Center, NAG 5-3249, 
and SAO contract SAO SV1-61010 to MIT.
 B.G.P. acknowledges the support of the VLBI staff at the U.S. Naval
 Observatory. We gratefully acknowledge the VSOP Project, which is led
 by the Japanese Institute of Space and Astronautical Science in
 cooperation with many organizations and radio telescopes around the
 world. The Australia Telescope is funded by the Commonwealth
Government for operation as a National Facility by the CSIRO.


\begin{thebibliography}{}
\bibitem[Chartas et al. (2000)]{cha00} Chartas, G. et al. 2000, \apj,
in press.
\bibitem[Elvis and Fabbiano (1984)]{elvis84} Elvis, M., and Fabbiano,
G. 1984, \apj, 280, 91
\bibitem[Fichtel et al. (1994)]{fic94} Fichtel, C. E. et al. 1994,
\apjs, 94, 551
\bibitem[Hirabayashi et al. (1998)]{hir98} Hirabayashi et al. 1998, Science, 281, 1825
\bibitem[Hunstead (1971)]{hun71} Hunstead, R.W. 1971, \mnras, 152, 277
\bibitem[Jerius et al. (2000)]{jerius00} Jerius, D., Edgar, R. J.,
Gaetz, T. J., McNamara, B. R., Schwartz, D. A., VanSpeybroeck, L. P.,
and Zhao, P. 2000, \procspie, 4012, in press.
\bibitem[Jones, O'Dell, and Stein (1972)]{jones74} Jones, T. W.,
O'Dell, S. L., and Stein, W. A. 1974, \apj, 188, 353
\bibitem[Lovell et al. (2000)]{lovell00} Lovell, J. E. J. 2000, in
Astrophysical Phenomena Revealed by Space VLBI, eds. H. Hirabayashi,
P. G. Edwards, and D. W. Murphy (Sagamihara:ISAS), 215
\bibitem[Madejski and Schwartz (1983)]{mad83} Madejski, G. M., and
Schwartz, D. A. 1984, \apj, 275, 467
\bibitem[Miley (1980)]{miley80} Miley, G 1980, \araa, 18, 165
\bibitem[Perez-Fournon (1985)]{perez85} Perez-Fournon, I. 1985, in
Active Galactic Nuclei, ed. J.E. Dyson (Manchester: Manchester
University Press), 300 
\bibitem[Piner and Kingham (1998)]{pin98} Piner, B. G., and Kingham,
K. A.  1998, \apj, 507, 706
\bibitem[Rees (1966)]{rees66} Rees, M. J. 1966, \nat, 211, 468
\bibitem[Savage, Browne, and Bolton (1976)]{sav76} Savage, A., Browne, I.W.A., and Bolton, J. G. 1976, \mnras, 177, 77P
\bibitem[Tingay et al. (1998)]{tin98} Tingay, S. J. et al. 1998, \apj,
497,594
\bibitem[Tingay et al. (2000)]{tin99} Tingay, S. J. et al. 2000,
Adv. Sp. Res., in press

\bibitem[Weisskopf et al. (2000)]{weiss00} Weisskopf, M. C.,
Tananbaum, H. D., VanSpeybroeck, L. P., and O'Dell, S. L. 2000,
\procspie, 4012, in press.
\bibitem[Yaqoob et al. (1998)]{yaqoob98} Yaqoob, T., George, I. M.,
Turner, T. J., Nandra, K., Ptak, A., and Serlemitsos, P. J. 1998,
\apjl, 505, L87
\bibitem[Wood et al. (1984)]{wood84} Wood, K. S., Meekins, J. F., Yentis, D. J.,
 Smathers, H. W., McNutt, D. P.,
 Bleach, R. D., Friedman, H.,
 Byram, E. T., Chubb, T. A., Meidav, M. 1984, \apjs, 56, 507
\end{thebibliography}
\end{document}